\documentclass[prd,twocolumn,amsmath,amssymb,nofootinbib,superscriptaddress]{revtex4-1}

\usepackage{subfigure}

\usepackage[usenames,dvipsnames,svgnames,table]{xcolor}
\usepackage{graphicx}
\usepackage{dcolumn}
\usepackage{bm}
\usepackage{multirow}
\usepackage{amsmath,amsfonts,amssymb}
\usepackage[colorlinks,pdfstartview=FitH]{hyperref}
\usepackage{siunitx}

\definecolor{red}{HTML}{FF0000}
\definecolor{blue}{HTML}{4169E1}
\definecolor{darkblue}{HTML}{00008B}
\definecolor{lightblue}{HTML}{ADD8E6}
\definecolor{lightpurple}{HTML}{C6AEC7}
\definecolor{magenta}{HTML}{E238EC}
\definecolor{cyan}{HTML}{00FFFF}

\hypersetup{linkcolor=blue,citecolor=magenta,filecolor=black,urlcolor=magenta}
\usepackage{multimedia}
\usepackage{natbib}
\usepackage{soul}

\DeclareMathOperator{\re}{Re}
\DeclareMathOperator{\im}{Im}
\begin{document}

\title{Functional renormalization group approach to the Yang-Lee edge singularity}

\author{X.\ An}
\affiliation{
  Department of Physics, University of Illinois at Chicago, Chicago, IL 60607, USA
}
\author{D.\ Mesterh\'azy}
\affiliation{
  Albert Einstein Center for Fundamental Physics, Institute for Theoretical Physics, University of Bern, 3012 Bern, Switzerland
}
\author{M.\ A.\ Stephanov}
\affiliation{
  Department of Physics, University of Illinois at Chicago, Chicago, IL 60607, USA
}

\date{\today}

\begin{abstract}
  We determine the scaling properties of the Yang-Lee edge singularity as described by a one-component scalar field theory with imaginary cubic coupling, using the nonperturbative functional renormalization group in $3 \leq d \leq 6$ Euclidean dimensions. We find very good agreement with high-temperature series data in $d = 3$ dimensions and compare our results to recent estimates of critical exponents obtained with the four-loop $\epsilon = 6-d$ expansion and the conformal bootstrap. The relevance of operator insertions at the corresponding fixed point of the RG $\beta$ functions is discussed and we estimate the error associated with $\mathcal{O}(\partial^4)$ truncations of the scale-dependent effective action.
\end{abstract}

\pacs{64.60.ae, 11.10.Gh}

\maketitle

\section{Introduction}

With the pioneering work of Yang and Lee a new perspective on the properties of statistical systems was established by pointing out the importance of the distribution of zeros of the partition function \cite{Yang:1952be,Lee:1952ig}. Expressed in terms of an external parameter, which we shall denote by $z$, the partition function $Z = Z(z)$ of a finite system can in general be expressed in terms of its roots $z_{\alpha}$ in the complex plane, i.e., we may write $Z = \prod_{\alpha} (z - z_{\alpha})$. Their significance appears in the thermodynamic limit, $V\to\infty$, when they coalesce along one-dimensional curves that separate different infinite volume behaviors of the partition function.\footnote{In principle, the zeros may accumulate on a dense set in parameter space, which must not necessarily be one dimensional. However, such a scenario is not relevant to this work.}  These curves can be viewed as cuts that distinguish different branches of the free energy (or grand canonical potential) $\Omega = - \beta \ln Z = - \beta V\int \textrm{d}\theta\hspace{1pt} g(\theta) \ln \left[z - z(\theta)\right]$, where $g(\theta)$ corresponds to the normalized density of zeros ($\int \textrm{d}\theta \hspace{1pt} g(\theta) = 1$) on a curve parametrized as $z(\theta)$ and $\beta = 1/T$ is the inverse temperature ($k_B = 1$). Clearly, once the location of the zeros, or cuts they coalesce into, $z(\theta)$, and the distribution $g(\theta)$ is known, in principle, all thermodynamic properties of the system can be calculated. This has led to numerous efforts to determine $g(\theta)$ for a wide range of lattice models via numerical methods \cite{Abe:1967,Suzuki:1967,Kortman:1971,Itzykson:1983gb} and also experimentally \cite{Binek:1998,Binek:2001,Peng:2015}. Besides providing a rigorous basis to study the thermodynamic properties of finite lattice systems, such attempts have also helped to elucidate features of fundamental theories. Drawing on the principle of universality they have led to important insights into the phase diagram of strongly-interacting matter at nonvanishing baryon densities \cite{Halasz:1996jg,Ejiri:2005ts,Stephanov:2006dn}.

Typically, for lattice spin models at temperature $T$ and external field $H$ the natural variable in terms of which the partition function is a polynomial is $z = \exp \left(- 2 \beta H \right)$. The zeros of $Z(z)$ are commonly referred to as Yang-Lee or Lee-Yang zeros. In particular, for the ferromagnetic Ising model one finds these zeros  distributed along the unit circle $z = \exp ( i \theta )$, where $\theta = 2 i \beta H$ and $H$ is imaginary. This has been proven rigorously and is known as the Yang-Lee circle theorem \cite{Lee:1952ig,Asano:1968a,*Asano:1968b,Kawabata:1968,Suzuki:1968,Griffiths:1969,Suzuki:1971,Ruelle:1973,Lebowitz:2012}. Depending on the temperature one may distinguish different scenarios: In the low-temperature region of the Ising model ($T < T_c$), the set of zeros crosses the positive real $z$-axis at $z = 1$ ($\theta = 0$), which indicates the presence of a first-order phase transition as one traverses the $\re H = 0$ axis from positive to negative real $H$ (or vice versa). On the other hand, in the high-temperature region ($T > T_c$) one observes a finite gap in the distribution $g(\theta) = 0$ for $|\theta| < \theta_g$ that closes as $T \rightarrow T_c^{+}$ \cite{Abe:1967,Suzuki:1967}. Thus, for $T > T_c$ the free energy is analytic along the real $H$ axis. However, at the edge of the gap $\theta = \pm\theta_g$, corresponding to \textit{imaginary} values of the magnetic field $H = \pm i |H_c(T)|$, the distribution of zeros exhibits nonanalytic behavior, i.e., $g(\theta) \simeq \left( |\theta| - \theta_g \right)^{\sigma}$, for $|\theta| \gtrsim \theta_g$, characterized by the exponent $\sigma$ \cite{Kortman:1971}. As pointed out by Fisher \cite{Fisher:1978pf} this behavior can be identified with a thermodynamic singularity that yields a divergence in the isothermal susceptibility $\chi = (\partial M / \partial H)_{T} \sim |H-H_c(T)|^{\sigma - 1}$, where $M$ is the magnetization. Thus, the Yang-Lee edge singularity at nonvanishing imaginary values of the field is similar to a conventional second order phase transition \cite{Fisher:1978pf,Kurtze:1979zz}.

In contrast to the well-known $\phi^4$ field theory that describes the critical point of the Ising model at $T = T_c$ and $H = 0$, the field theory at the Yang-Lee edge point, the $\phi^3$ theory, admits no discrete reflection symmetry and is therefore characterized by only one independent (relevant) exponent. In two dimensions the corresponding universality class has been identified with that of the simplest nonunitary conformal field theory (CFT), the  minimal model $M_{2,5}$, with central charge $c = -22/5$ \cite{Cardy:1985yy}. This allowed to exploit conformal symmetry in two dimensions to calculate the scaling exponent $\sigma(d=2) = -1/6$, which has been confirmed with remarkable accuracy by series expansions \cite{Baker:1986,Kim:2006}, as well as by comparing with experimental high-field magnetization data \cite{Binek:1998,Binek:2001}. Furthermore, using integral kernel techniques it is possible to establish the exact result $\sigma(d=1) = -1/2$ \cite{Fisher:1978pf,Kurtze:1979zz}. On the other hand, most of our knowledge in the region $2 < d < 6$ relies on appropriately resummed results from the $\epsilon = 6 - d$ expansion \cite{deAlcantaraBonfim:1980pe,deAlcantaraBonfim:1981sy,Gracey:2015tta}, strong-coupling expansions \cite{Butera:2012tq}, Monte Carlo methods \cite{Lai:1995,Hsu:2005}, and conformal bootstrap \cite{Gliozzi:2014jsa}. Note that in contrast to the Ising critical point (described by $\phi^4$ theory), the upper critical dimension of the Yang-Lee edge point (described by $\phi^3$ theory) is $d_c = 6$ and therefore, fluctuations are important even above dimension $d = 4$.

Recently, there has been renewed interest in the Yang-Lee edge point for which the renormalization group (RG) $\beta$ functions to four-loop order in the $\epsilon$ expansion were determined in Ref.\ \cite{Gracey:2015tta} and the corresponding critical exponents (obtained from constrained Pad\'e approximants) were compared to estimates from other methods. In light of these developments, we examine the critical scaling properties of the Yang-Lee edge with the nonperturbative functional RG \cite{Wetterich:1989xg,Wetterich:1992yh} for dimensions $3 \leq d \leq 6$. In contrast to the $\epsilon$ expansion, the functional RG does not rely on the expansion in a small parameter and is therefore ideally suited to investigate the critical behavior of the Yang-Lee edge away from $d_c = 6$. However, care must be taken to address possible systematic errors that arise from the truncation of the infinite hierarchy of flow equations. We show that that these errors are under control and comment on the quality of different truncations. In summary, we find that the obtained values for the critical exponents are in good agreement with previous results obtained in $d = 3$ dimensions using high-temperature series expansions \cite{Kurtze:1979zz}, the three- and four-loop $\epsilon$ expansion around $d = 6$ \cite{deAlcantaraBonfim:1980pe,deAlcantaraBonfim:1981sy} as well as other methods \cite{Butera:2012tq,Lai:1995,Hsu:2005,Gliozzi:2014jsa}. We observe that derivative interactions have an important effect on the stability of the scaling solution and need to be taken into account properly in the framework of the nonperturbative functional RG.

The outline of this article is as follows: First, in Sec.\ \ref{Sec:Nonperturbative functional RG}, we give an overview of the nonperturbative functional RG and the truncations employed in this work. In Sec.\ \ref{Sec:Critical equation of state and mean-field scaling prediction} we discuss the scaling properties of the critical equation of state and the mean-field theory at the Yang-Lee edge singularity. In Sec.\ \ref{Sec:Solving the RG flow equations} we consider the general properties of RG flow trajectories and in particular their infrared (IR) behavior. In Secs.\ \ref{Sec:Relevance of composite operators and quality of finite truncations} -- \ref{Sec:Residual regulator dependence and principle of minimal sensitivity} we summarize our results for the critical exponents at the Yang-Lee edge singularity and analyze the expected systematic errors for the truncations employed in this work. We close by comparing our estimates for the critical exponents to recent data from Refs.\ \cite{Gracey:2015tta} and \cite{Gliozzi:2014jsa} and conclude with an outlook on future work.

\section{Nonperturbative functional RG}
\label{Sec:Nonperturbative functional RG}

In this work, we employ a RG scheme that relies on a truncation of a hierarchy of flow equations derived from an exact flow equation for the scale-dependent effective action $\Gamma_k$ \cite{Wetterich:1989xg,Wetterich:1992yh}, i.e., the generating functional of one-particle irreducible (1PI) diagrams (for reviews see, e.g., Refs.\ \cite{Bagnuls:2000ae,Berges:2000ew,Polonyi:2001se,Pawlowski:2005xe,Delamotte:2007pf}), where $k$ denotes the RG scale parameter. The scale-dependent effective action is obtained from the functional Legendre transform
\begin{equation}
  \Gamma_k = \sup_J \left( \int \textrm{d}^dx \hspace{1pt} J(x) \phi(x) - W_k \right) - \Delta_k S ,
\end{equation}
of the scale-dependent generating functional of connected correlation functions
\begin{equation}
  W_k = \ln \int [d\varphi] \hspace{1pt} \exp\bigg\{ - S - \Delta_k S + \int \textrm{d}^dx \, J(x) \varphi(x)\bigg\} ,
  \label{Eq:GeneratingFunctional}
\end{equation}
with respect to the external source $J = J(x)$; $\phi = \delta W_k / \delta J$ is the scalar field expectation value. Here, we consider a classical action $S$ of a single-component scalar field
\begin{equation}
  S = \int \textrm{d}^{d}x \, \left\{ \frac{1}{2} ( \partial \varphi )^2 + U_{\Lambda}(\varphi) \right\} ,
\label{Eq:ClassicalAction}
\end{equation}
and the classical potential $U_{\Lambda}$ is specified in Sec.\ \ref{Sec:Critical equation of state and mean-field scaling prediction}.
The additional term $\Delta_k S$ in Eq.\ \eqref{Eq:GeneratingFunctional} is a quadratic functional 
\begin{equation}
  \Delta_k S = \frac{1}{2} \int \textrm{d}^dx \, \textrm{d}^dy \, \varphi(x) R_k(x,y) \varphi(y) ,
\end{equation}
and serves to regularize the theory in the IR; in particular, the regulator function $R_k(x,y) = R_k(-\Box_x) \delta^{(d)}(x-y)$, where $\Box \equiv \partial_{\mu}\partial_{\mu}$, is chosen in such a way that it leads to a decoupling of IR modes. We require that $\lim_{k\rightarrow 0} R_k = 0$ and  $\lim_{\Lambda\rightarrow \infty} R_{k = \Lambda} = \infty$, where $\Lambda$ is a characteristic scale that regularizes the theory in the ultraviolet (UV) and can formally be sent to infinity. In effect, this defines a one-parameter family of theories ($0 \leq k \leq \Lambda$), which interpolates between the classical action, $S = \lim_{k\rightarrow \Lambda} \Gamma_k$, and the full 1PI effective action, $\Gamma = \lim_{k\rightarrow 0} \Gamma_k$. Thus, the scale-dependent regulator function $R_k$ induces a functional RG flow
\begin{equation}
  \frac{\partial}{\partial s} \Gamma_k = \frac{1}{2} \int \! \frac{\textrm{d}^dq}{(2\pi)^d} \frac{\partial R_k(q)}{\partial s} \left[ \Gamma_k^{(2)}(\phi; q) + R_k(q) \right]^{-1} ,
  \label{Eq:FunctionalFlow}
\end{equation}
between these two limits, where $s = \ln (k/\Lambda)$ is a dimensionless scale parameter, and $\delta^{(d)}\big(\sum_{i = 1}^n p_i\big) \Gamma_k^{(n)}(\phi; p_1, p_2, \ldots , p_{n-1}) \equiv (2\pi)^{(n-1) d} {\delta^n \Gamma_k[\phi]}/{\delta \phi(p_1) \delta \phi(p_2) \cdots \delta\phi(p_n)}$. In principle, we may choose any (sufficiently smooth) regulator that satisfies the above limiting properties. For details of our implementation and necessary requirements imposed on the regulator function see Secs.\ \ref{Sec:Solving the RG flow equations} -- \ref{Sec:Residual regulator dependence and principle of minimal sensitivity}.

\begin{table}[!t]
  \setlength{\tabcolsep}{2pt}
  \def\arraystretch{1.75}
  \centering
  \begin{tabular}{lclccl} 
  Operator && Coupling &&& Canonical dimension \\
  \hline\hline
  $\delta\phi^n$ && $\bar{U}^{(n)}$ &&& $ \dim \bar{U}^{(n)} = d - n (d-2)/2$ \\
  $\delta\phi^n (\partial \phi)^2$ && $\bar{Z}^{(n)}$ &&& $\dim \bar{Z}^{(n)} = - n (d-2)/2$ \\
  $\delta\phi^n \left(\Box \phi\right)^2$ && $\bar{W}_1^{(n)}$ &&& $\dim \bar{W}_1^{(n)} = -2 - n (d-2)/2$ \\
  $\delta\phi^n (\partial \phi)^{2} \,\Box \phi$ && $\bar{W}_2^{(n)}$ &&& $\dim \bar{W}_2^{(n)} = -[d + 2 + n (d-2) ]/2$ \\
  $\delta\phi^n \big[ ( \partial \phi )^{2} \big]^2$ && $\bar{W}_3^{(n)}$ &&& $\dim \bar{W}_3^{(n)} = -d - n (d-2)/2$ \\
  \hline
  \hline
\end{tabular}
  \caption{\label{Tab:CanonicalDimensions}Operators and canonical dimension of associated parameters and couplings that appear in the expansion of $\Gamma_k$ [cf.\ Eq.\ \eqref{Eq:Truncation}]. Note that we drop the RG scale index $k$, since the canonical dimensions are defined at the Gaussian fixed point of the RG $\beta$ functions.}
\end{table}

Clearly, an exact solution for the full functional flow is not feasible in practice, so one has to rely on suitable approximations of Eq.\ \eqref{Eq:FunctionalFlow}. Here, we comment on the nature of our truncation and discuss its limitations. We use a truncated expansion in derivatives for the scale-dependent effective action \cite{Morris:1993qb,Morris:1994ie}
\begin{eqnarray}
  && \hspace{-8pt} \Gamma_k = \int \textrm{d}^dx \hspace{2pt} \bigg\{ U_k(\phi) + \frac{1}{2} Z_k(\phi) (\partial \phi)^{2} + \frac{1}{2} W_{1,k}(\phi) ( \Box\hspace{1pt} \phi )^{2}  \nonumber\\
  && \hspace{16pt} +\: \frac{1}{2} W_{2,k}(\phi) (\partial \phi)^{2} \,\Box\hspace{1pt} \phi + \frac{1}{2} W_{3,k}(\phi) \big[ ( \partial \phi )^{2} \big]^2 \bigg\} ,
\label{Eq:Truncation}
\end{eqnarray}
where $U_k$ is the scale-dependent effective potential, and the scale-dependent functions $Z_k$ and $W_{a,k}$, $a=1,2,3$, parametrize the contributions to order $\partial^4$ (up to total derivative terms). Furthermore, for each of these functions, we employ a finite series expansion in the fluctuation $\delta\phi_k = \phi - \bar{\phi}_k$ around a field configuration $\bar{\phi}_k$, which is assumed to be homogeneous in space [cf.\ Sec.\ \ref{Sec:Solving the RG flow equations}]. In effect, this corresponds to an \textit{ansatz} for $\Gamma_k$ that includes only a finite set of independent operators, each of which is parametrized by a single parameter or coupling that is field independent, e.g., $Z_k(\phi) (\partial \phi)^2 = \big( \bar{Z}_k^{(0)} +  \bar{Z}_k^{(1)} \delta\phi_k + \ldots \big) (\partial \phi)^2$, and $\bar{Z}_k^{(n)}\equiv Z_k^{(n)}(\bar{\phi}_k)$, $n \in \mathbb{N}$, and similar expansions apply to $U_k$ and $W_{a,k}$. The  canonical dimensions of these parameters are displayed in \mbox{Tab.\ \ref{Tab:CanonicalDimensions}}. Clearly, above dimension $d = 2$, $\bar{Z}_k^{(n)}$ and $\bar{W}_{a,k}^{(n)}$ are irrelevant as far as a counting of canonical dimensions goes, but this is not sufficient to conclude that this is also the case at a nontrivial (i.e., non-Gaussian) fixed point of the RG $\beta$ functions. Indeed, one of the objectives of this paper is to investigate their effect at the Yang-Lee edge point as well as on RG trajectories that approach this scaling solution in the IR. We should point out that similar truncations of the scale-dependent effective action were considered also in Refs.\ \cite{Wetterich:1991be,Seide:1998ir,Canet:2003qd,Litim:2010tt} to establish the critical exponents at the Ising critical point. Here, we study the scaling properties of Eq.\ \eqref{Eq:Truncation} in the presence of a nonvanishing external field, when the discrete reflection symmetry $\phi \leftrightarrow -\phi$ of the Ising model is explicitly broken and the system is tuned to the Yang-Lee edge critical point.

The flow equations for $U_k$, $Z_k$, and $W_{a,k}$, $a = 1,2,3$, are derived from the exact functional flow equation for $\Gamma_k$ [cf.\ Eq.\ \eqref{Eq:FunctionalFlow}] by applying functional derivatives and projecting them onto the appropriate momentum contributions, i.e.,
\begin{subequations}
\begin{eqnarray}
  && \hspace{-8.2pt} \frac{\partial}{\partial s} U_k = \frac{\partial}{\partial s} \left. \Gamma_k[\phi] \right|_{\phi=\textrm{const.}}, \label{Eq:ProjectionU} \\ 
  && \hspace{-8pt} \frac{\partial}{\partial s} Z_k = \lim_{p\rightarrow 0} \frac{\partial}{\partial p^2} \frac{\partial}{\partial s} \Gamma_k^{(2)}(\phi; p) , \label{Eq:ProjectionZ} \\
  && \hspace{-17pt} \frac{\partial}{\partial s} W_{1,k} = \lim_{p\rightarrow 0} \frac{\partial}{\partial (p^2)^2} \frac{\partial}{\partial s} \Gamma_k^{(2)}(\phi; p) , \label{Eq:ProjectionW1} \\
  && \hspace{-17pt} \frac{\partial}{\partial s} W_{2,k} = \frac{1}{2} \lim_{p_i\rightarrow 0} \frac{\partial}{\partial (p_1 \hspace{-2pt}\cdot \hspace{-1pt} p_2)^2} \frac{\partial}{\partial s} \Gamma_k^{(3)}(\phi; p_1, p_2) , \label{Eq:ProjectionW2} \\
  && \hspace{-16.5pt} \frac{\partial}{\partial s} W_{3,k} = -\frac{1}{4} \lim_{p_i\rightarrow 0} \left[\frac{\partial}{\partial (p_2 \hspace{-2pt}\cdot \hspace{-1pt} p_3)} - \frac{1}{2}\frac{\partial}{\partial (p_1 \hspace{-2pt}\cdot \hspace{-1pt} p_2)} \right. \nonumber\\ && \hspace{30pt} \left. -\: \frac{1}{2}\frac{\partial}{\partial (p_1\hspace{-2pt}\cdot \hspace{-1pt}p_3)} \right] \frac{\partial}{\partial p_1^2} \frac{\partial}{\partial s} \Gamma_k^{(4)}(\phi; p_1, p_2, p_3) , \label{Eq:ProjectionW3}
\end{eqnarray}
\end{subequations}
where $p\cdot q \equiv p_{\mu}q_{\mu}$. The corresponding RG flow equations for the field-independent parameters $\bar{Z}_k^{(n)}$ and $\bar{W}_{a,k}^{(n)}$ can be derived from Eqs.\ \eqref{Eq:ProjectionU} -- \eqref{Eq:ProjectionW3} by suitable differentiation and successive projection onto the reference field configuration $\bar{\phi}_k$ that enters the series expansion. We do not display them at this point but refer the reader to supplementary material available online \cite{Mesterhazy:2016}.

The RG flow equations display the following chain of dependencies
\begin{equation}
  U_k \leftarrow \{ Z_k , W_{1,k} \} \leftarrow \{ W_{2,k} , W_{3,k} \} \leftarrow \ldots , 
  \label{Eq:ChainDependencies}
\end{equation}
where the ellipsis denotes higher order contributions that we have chosen to neglect in our \textit{ansatz}, Eq.\ \eqref{Eq:Truncation}. That is, the RG flow equation for the scale-dependent effective potential $U_k$ depends on the quantities $Z_k$ and $W_{1,k}$, but is independent of $W_{2,k}$ and $W_{3,k}$ etc. We exploit this structure explicitly by truncating the hierarchy Eq.\ \eqref{Eq:ChainDependencies} at the second level, i.e., we set $W_{2,k} = W_{3,k} = 0$ in Eq.\ \eqref{Eq:Truncation}, while $U_k$, $Z_k$, and $W_k\equiv W_{1,k}$ are expanded to some finite order in $\delta\phi_k$. Note that the order of the employed expansion might be different for each of these coefficients. Similar approximations have led to reasonable estimates of the critical scaling exponents at the Ising critical point \cite{Canet:2003qd,Litim:2010tt} and we expect that this is also the case for the Yang-Lee edge critical point.

\section{Critical equation of state and mean-field scaling prediction}
\label{Sec:Critical equation of state and mean-field scaling prediction}

Here, we consider a classical potential of the following form 
\begin{equation}
  U_{\Lambda} = \frac{1}{2} t_{\Lambda} \varphi^2 + \frac{1}{4!} \lambda_{\Lambda} \varphi^4 + h_{\Lambda} \varphi,
  \label{Eq:ClassicalPotential}
\end{equation}
with a nonvanishing coupling to a symmetry-breaking field $h_{\Lambda}$, and $t_{\Lambda} \sim T - T_c$, with $T_c$ the critical temperature at the Ising critical point. Upon integration of the RG flow equations \eqref{Eq:ProjectionU} -- \eqref{Eq:ProjectionW1} down from the cutoff scale $\Lambda$ to the IR, the parameters and couplings of the classical potential acquire a scale dependence. In fact, the corresponding scale-dependent effective potential $U_k$ for $0\leq k < \Lambda$ will typically include a large number of fluctuation-induced interactions. The full effective potential is obtained only when the scale parameter $k$ is sent to zero and all modes have been integrated out, i.e., $U = \lim_{k\rightarrow 0} U_k$.

In order to arrive at a critical point in the IR the relevant parameters of the classical action need to be tuned to their respective critical values, while all other parameters or couplings are kept constant. That is, in the case of the Yang-Lee edge critical point, we fix $\lambda_{\Lambda}$, $|h_{\Lambda}| > 0$, and tune $t_{\Lambda}$ to its critical value $t_{\Lambda,c} = t_{\Lambda,c}(h_{\Lambda}) > 0$, for which $\bar{U}^{(1)} \equiv \lim_{k\rightarrow 0} \bar{U}_k^{(1)} = 0$ and $\bar{U}^{(2)} \equiv \lim_{k\rightarrow 0} \bar{U}_k^{(2)} = 0$ in the IR limit. At the Yang-Lee edge critical point, the first and second derivative are evaluated at a nonvanishing, \textit{imaginary} field expectation value $\bar{\phi}$. In the critical domain, the equation of state satisfies the scaling form
\begin{equation}
  U'(\phi) = \delta\phi |\delta\phi|^{\delta-1} f\left(\delta t_{\Lambda} |\delta\phi|^{-1/\beta}\right) ,
  \label{Eq:ScalingForm}
\end{equation}
where $\delta\phi = \phi - \bar{\phi}$ and $f = f(x)$ is a universal, dimensionless scaling function, which is uniquely defined up to normalization. The critical exponents $\beta$ and $\delta$ characterize the asymptotic scaling behavior of the (residual) magnetization $\delta\phi$ for vanishing $U'(\phi) = \delta h$ and $\delta t_{\Lambda} = t_{\Lambda} - t_{\Lambda,c}$, respectively. Here, the parameter $\delta h \sim H - H_c$, measures the deviation from the critical field strength $H_c = \pm i |H_c(T)|$, and $T > T_c$ for the range of values of $\delta t_{\Lambda}$ studied in this work.

Before we go on to consider the solution of the RG flow equations \eqref{Eq:ProjectionU} -- \eqref{Eq:ProjectionW1}, we discuss the mean-field scaling prediction. Since there is no scale dependence in this case, we simply drop the $k$ (or $\Lambda$) index on all parameters. It is useful to express the potential in terms of an expansion in field differences $\delta\varphi = \varphi - \bar{\varphi}$ around a reference field configuration $\bar{\varphi}$, which is defined such that $U'(\bar{\varphi}) = 0$. According to the strategy outlined above, we fix $|h| > 0$ and inquire about possible critical points, by imposing in addition the condition that $U''(\bar{\varphi}) = 0$. We derive two independent scaling solutions, which we identify as
\begin{equation}
  t_c = \lambda/2 \left( \pm i 3 h/\lambda \right)^{2/3} . 
\end{equation}
Assuming that $t_c > 0$ we see that the corresponding critical field value $h_c = \pm i \lambda/3  \left( 2 t_c / \lambda\right)^{3/2}$ is imaginary in accordance with the Yang-Lee theorem \cite{Yang:1952be, Lee:1952ig}. Near the critical point $U'(\varphi)$ satisfies the scaling form \eqref{Eq:ScalingForm} with $\delta = 2$ and $\beta = 1$. Other critical exponents that characterize the power-law singularities of various thermodynamic quantities can be determined via scaling relations \cite{Pelissetto:2002}. That is, in the absence of fluctuations the anomalous dimension vanishes, $\eta = 0$, and we obtain the following scaling exponents: $\alpha = -1$, $\gamma = 1$, $\nu = 1/2$, and $\nu_c = 1/4$. Note that the exponent $\alpha$ is negative and therefore, at the mean-field level, the specific heat does not diverge at the Yang-Lee edge point.

\section{Solving the RG flow equations}
\label{Sec:Solving the RG flow equations}

\begin{figure*}[!t]
  \includegraphics[width=0.41\textwidth]{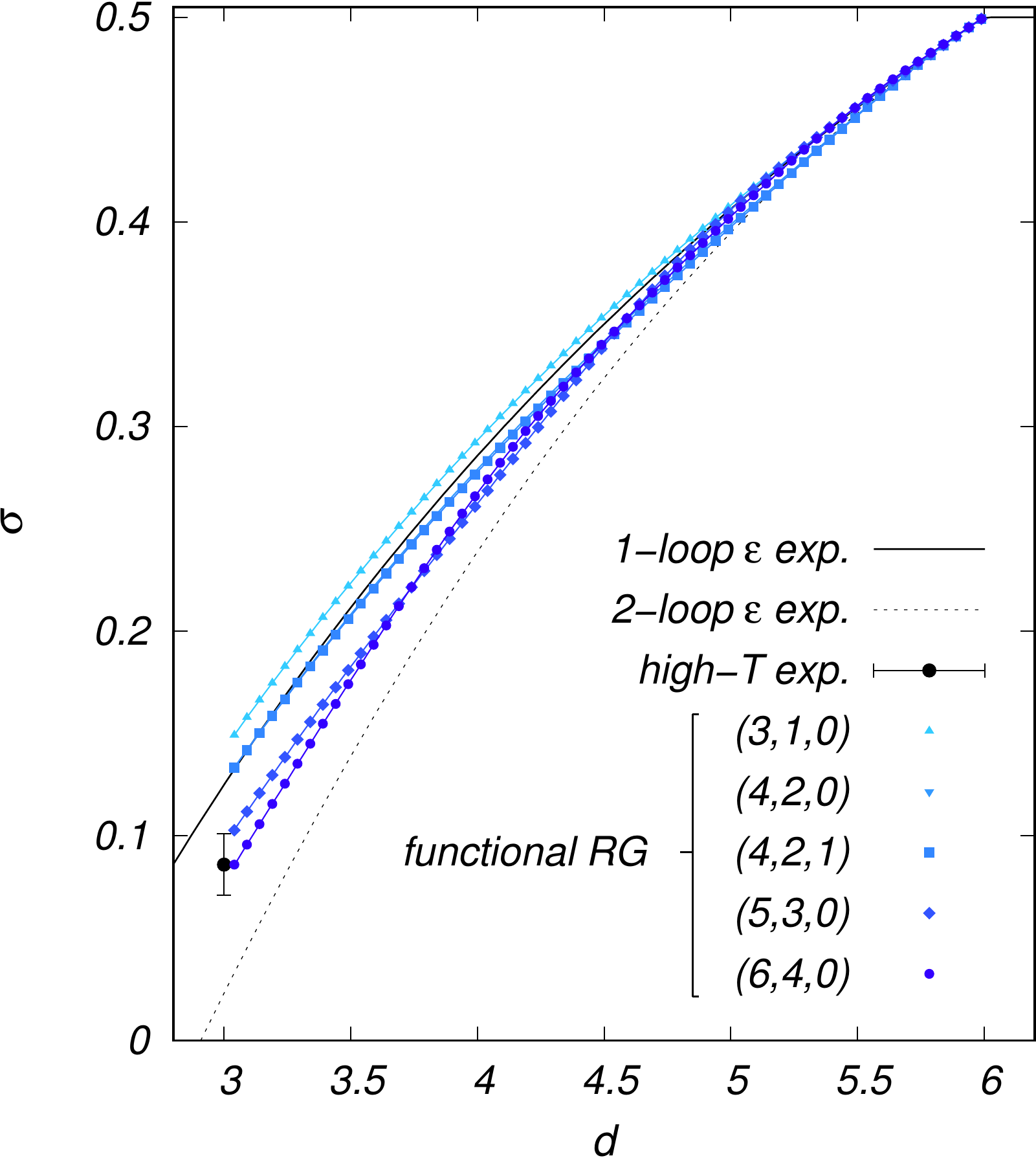} \hspace{30pt}
  \includegraphics[width=0.42\textwidth]{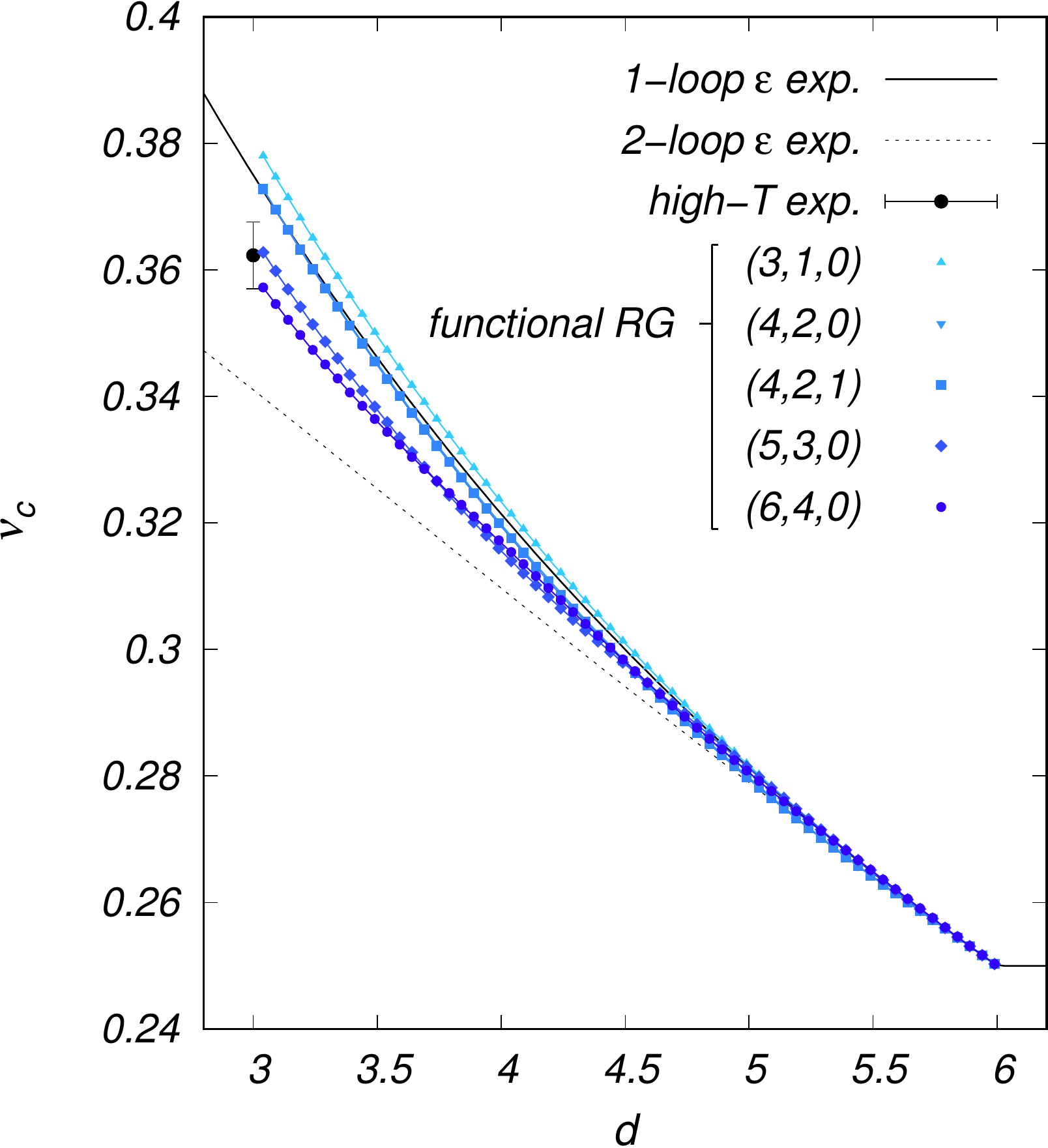}
  \caption{\label{Fig:CriticalExponents}Critical exponents $\sigma = 1/\delta$ and $\nu_c$ as a function of Euclidean dimension $d$ for different truncations of the scale-dependent effective action $\Gamma_k$, as specified by the set of integers $(n_U, n_Z, n_W)$ [cf.\ Sec.\ \ref{Sec:Solving the RG flow equations}]. The data for the truncation $(4,2,1)$ lies almost exactly on top of that for $(4,2,0)$. Shown in comparison are results from the one- and two-loop $\epsilon$ expansion \cite{Macfarlane:1974vp,Fisher:1978pf} as well as high-temperature series expansion data ($d = 3$) \cite{Fisher:1978pf,Kurtze:1979zz}. We observe that the numerical accuracy of the functional RG results improves significantly as one goes to higher orders in the derivative and field expansion, respectively.}
\end{figure*}

To solve the RG equations we specify the classical action $S = \int \textrm{d}^{d}x \, \left\{ \frac{1}{2} (\partial \varphi)^2 + U_{\Lambda}(\varphi) \right\}$, which is defined in terms of the short-distance potential $U_{\Lambda}$, and integrate the flow equations down to $s \rightarrow -\infty$. The classical potential is given in Eq.\ \eqref{Eq:ClassicalPotential} and the coefficients that parametrize the kinetic contribution to the action are $Z_{\Lambda} = 1$ and $W_{\Lambda} = 0$.

We use a truncated series expansion for the scale-dependent effective potential $U_k$ as well as for the field-dependent renormalization factors $Z_k$ and $W_k$ ($0\leq k \leq \Lambda$). Such a strategy is often sufficient to extract the leading or subleading critical scaling behavior \cite{Litim:2002cf,Litim:2003kf,Canet:2003qd,Bervillier:2007rc,Litim:2010tt}. The employed expansion is organized around a nonvanishing, imaginary, and homogeneous field configuration $\bar{\phi}_k$, which depends on the scale parameter $k$, and is defined in the following way: 1) At the cutoff scale $\Lambda$, $\bar{\phi}_{k = \Lambda} = \bar{\varphi}_{\Lambda}$ is a solution to $U_{\Lambda}''(\bar{\varphi}_{\Lambda}) = \tau$, and 2) the scale derivative of $\bar{U}_k^{(2)} \equiv U_k''(\bar{\phi}_k)$, evaluated at $\bar{\phi}_k = \bar{\phi} + \bar{\chi}_k$, satisfies
\begin{eqnarray}
  \frac{\textrm{d}}{\textrm{d}s} \bar{U}_k^{(2)} &=& \frac{\partial}{\partial s} \bar{U}_k^{(2)} + \bar{U}_k^{(3)} \frac{\textrm{d}\bar{\chi}_k}{\textrm{d}s} = 0 .
  \label{Eq:Constraint}
\end{eqnarray}
Of course, the imaginary field expectation value $\bar{\phi}$ is scale independent and therefore $\textrm{d} \bar{\phi}_k / \textrm{d}s = \textrm{d}\bar{\chi}_k / \textrm{d}s$. Note that $\lim_{k\rightarrow 0} \bar{\chi}_k = 0$, i.e., $\lim_{k\rightarrow 0} \bar{\phi}_k = \bar{\phi}$, only when $\tau = 0$ and the system has been tuned to criticality. Clearly, conditions 1) and 2) fix one parameter of the model $\bar{U}_k^{(2)} = \tau$, at the expense of introducing another scale-dependent quantity, the field configuration $\bar{\chi}_k$, for which we obtain
\begin{equation}
  \frac{\textrm{d}\bar{\chi}_k}{\textrm{d}s} = - \big(\bar{U}_k^{(3)}\big)^{-1} \frac{\partial}{\partial s} \bar{U}_k^{(2)} .
\end{equation}
Note that the corresponding set of flow equations requires that $|\bar{U}_k^{(3)}| > 0$ for all $0\leq k \leq \Lambda$. This does not hold true in the vicinity of the Ising critical point and therefore, the chosen expansion point is not adequate to investigate the scaling properties for critical points on the $\phi \leftrightarrow -\phi$ symmetry axis ($H = 0$).

Eq.\ \eqref{Eq:Constraint} fixes the second derivative of the scale-dependent effective potential at all scales and therefore the expansion of the scale-dependent effective potential reads
\begin{equation}
  U_k = \bar{U}_k^{(0)} + \bar{U}_k^{(1)} \delta\phi_k + \frac{1}{2} \tau \hspace{1pt} \delta\phi_k^2 + \sum_{n = 3}^{n_{U}} \frac{1}{n!} \bar{U}_k^{(n)} \delta\phi_k^n .
  \label{Eq:PotentialSeries}
\end{equation}
Here, the sum runs up to some finite integer value $n_U$, which defines our truncation for the scale-dependent effective potential with the prescribed expansion point. The coefficients $\bar{U}_k^{(n)}$, $n\in \mathbb{N}$, are related to the couplings and parameters of the classical potential at the short-distance cutoff $\Lambda$, i.e., $\bar{U}_{\Lambda}^{(0)} = \bar{\varphi}_{\Lambda} \left[h_{\Lambda} + 1/12 \hspace{2pt} (5 t_{\Lambda} + \tau) \bar{\varphi}_{\Lambda}\right]$, $\bar{U}_{\Lambda}^{(1)} = h_{\Lambda} + (2 t_{\Lambda} + \tau) / 3 \hspace{2pt} \bar{\varphi}_{\Lambda}$, and $\bar{U}_{\Lambda}^{(3)} = \lambda_{\Lambda}/6 \hspace{2pt}\bar{\varphi}_{\Lambda}$, $\bar{U}_{\Lambda}^{(4)} = \lambda_{\Lambda}$, while $\bar{U}_{\Lambda}^{(n)} = 0$, for $n > 4$. Similarly, the expansions for $Z_k$ and $W_k$ read
\begin{subequations}
\begin{eqnarray}
  Z_k &=& \sum_{n = 0}^{n_Z-1} \frac{1}{n!} \bar{Z}_k^{(n)} \delta\phi_k^n , \\
  W_k &=& \sum_{n = 0}^{n_W-1} \frac{1}{n!} \bar{W}_k^{(n)}\delta\phi_k^n ,
  \label{Eq:RenormalizationFactorSeries}
\end{eqnarray}
\end{subequations}
with $\bar{Z}_{\Lambda}^{(0)} = 1$, $\bar{Z}_{\Lambda}^{(n)} = 0$ for $n > 0$, and $\bar{W}_{\Lambda}^{(n)} = 0$ for $n\in \mathbb{N}$. We define $Z_k \equiv 0$ if $n_Z = 0$ and $W_k \equiv 0$ if $n_W = 0$. In the following, we denote these type of series truncations in short by the set of integers $(n_U, n_Z, n_W)$. $n_U$ is considered as a free parameter, while $n_Z$ and $n_W$ are chosen such that $\max_{n_Z} \dim \bar{Z}_k^{(n_Z)} \leq \dim \bar{U}_k^{(n_U)}$ and $\max_{n_W} \dim \bar{W}_k^{(n_W)} \leq \dim \bar{U}_k^{(n_U)}$ in $d = 6$ dimensions. This choice defines what we consider to be consistent truncations (see Sec.\ \ref{Sec:Relevance of composite operators and quality of finite truncations}).

\newpage
Substituting Eqs.\ \eqref{Eq:PotentialSeries} -- \eqref{Eq:RenormalizationFactorSeries} back into \eqref{Eq:ProjectionU} -- \eqref{Eq:ProjectionW1} we obtain a finite set of flow equations for the coefficients of the series expansion. In this work, we consider expansions of order up to $(n_U, n_Z, n_W) = (7, 5, 0)$ and $(5, 3, 2)$, which yields a coupled set of partial differential equations of up to $12$ and $10$ parameters, respectively. The Yang-Lee scaling solution is identified by inspecting the behavior of the first and second derivatives of the effective potential, which should satisfy $\bar{U}^{(1)} = \bar{U}^{(2)} = 0$, while $\im \bar{U}^{(2n)} = \re \bar{U}^{(2n+1)} = 0$, for $n\in \mathbb{N}$. Note that all of these coefficients are defined at a reference field configuration $\bar{\phi}$ (where $\lim_{k\rightarrow 0} \bar{\chi}_k = 0$), which is imaginary, corresponding to the imaginary magnetic field $H_c = \pm i |H_c(T)|$, with $T > T_c$.

\vskip 15pt
We introduce the following short-hand notation for the renormalization factor $\bar{Z}_k \equiv Z_k^{(0)}(\bar{\phi}_{k})$, which satisfies $\bar{Z}_k \sim (k/\Lambda)^{-\eta}$ at the critical point. Starting from a set of initial values for the parameters and couplings in the classical action, which are tuned to their critical values, we may therefore define the anomalous dimension by the corresponding value in the IR:
\begin{equation}
  \eta = - \lim_{k\rightarrow 0} \frac{\partial}{\partial s} \ln \bar{Z}_k . 
\end{equation}
Note that the anomalous dimension at the Yang-Lee edge critical point is negative for all values of $1 \leq d < 6$.

\section{Critical scaling exponents and hyperscaling relations} 
\label{Sec:Critical scaling exponents and hyperscaling relations}

The critical exponents at the Yang-Lee edge critical point are extracted by a stability analysis of the scaling solution with respect to perturbations with those operators included in our \textit{ansatz} Eq.\ \eqref{Eq:Truncation}. That is, for any finite truncation of the scale-dependent effective action, we obtain a finite set critical exponents corresponding to the eigenvalues $\lambda_n$ of the stability matrix,
\begin{equation}
  \gamma_{mn} = \frac{\partial \beta_m\big(\{ \bar{g}_{{}_{\ast}, l} \}_{l\in I}\big)}{\partial \bar{g}_{n,k}} ,
\end{equation}
which is evaluated at the fixed point of the RG $\beta$ functions, $\beta_m \equiv \partial \bar{g}_{m,k} / \partial s$, i.e.,
\begin{equation}
  \beta_m\big(\{ \bar{g}_{{}_{\ast}, n}\}_{n\in I}\big) = 0 .
\end{equation}
The $\beta$ functions are derived for the dimensionless, renormalized parameters and couplings of the model, $\bar{g}_{n,k}$, $n\in I = \{1, 2, \ldots, n_U + n_Z + n_W\}$, which are given by $\bar{g}_{1,k} = k^{-(d+2)/2} \bar{Z}_k^{-1/2} \bar{U}_k^{(1)}$, $\bar{g}_{2,k} = k^{(2-d)/2} \bar{Z}_k^{1/2} \bar{\chi}_k$ , etc. We order the eigenvalues $\lambda_n$, $n = 1,2, \ldots$, according to their values in $d = 6$ dimensions, where they are identical to the canonical dimension of the parameters and couplings associated with the operators that appear in $\Gamma_k$, e.g., $\lambda_1(d = 6) = \dim \bar{U}^{(1)} \geq \lambda_2(d = 6) = \dim \bar{\chi} \geq \ldots$. Of course, as the eigenvalues are analytically continued to dimensions below $d = 6$, this ordering might change.

We observe that the largest eigenvalue $\lambda_1 \equiv 1/\nu_c$ satisfies the following scaling relation
\begin{equation}
  1/\nu_c = ( d + 2 - \eta ) / 2 ,
  \label{Eq:NuC}
\end{equation}
and therefore, the critical exponent $\nu_c$ is determined completely in terms of the anomalous dimension $\eta$.
The Yang-Lee edge critical point is known to exhibit another hyperscaling relation, which follows from the equation of motion of the $\phi^3$ theory \cite{Amit:1984ms} and can be written as
\begin{equation}
  \lambda_1 + \lambda_2 = d ,
  \label{Eq:HyperScaling}
\end{equation}
with $\lambda_2\equiv 1/\nu$, from which we obtain
\begin{equation}
  1/\nu = ( d - 2 + \eta ) / 2 .
  \label{Eq:Nu}
\end{equation}
Furthermore, from scaling and hyperscaling relations, one can derive
\begin{equation}
  \sigma = \frac{1}{\delta} = \frac{d - 2 + \eta}{d + 2 - \eta} ,
  \label{Eq:Sigma}
\end{equation}
and $\beta = 1$, independent of dimension \cite{deAlcantaraBonfim:1980pe}. Note, however, that for any finite truncation of $\Gamma_k$ scaling relations between critical exponents need not necessarily be satisfied and therefore should be checked explicitly. This applies to both Eq.\ \eqref{Eq:Nu} and to Eq.\ \eqref{Eq:Sigma}. Taking Eq.\ \eqref{Eq:Nu} for example, one may define the relative difference $\Delta \lambda_2 / [(d - 2 + \eta)/2] = 2 \lambda_2/ (d - 2 + \eta) - 1$ as an indicator for the quality of the employed truncation at the Yang-Lee edge fixed point. We observe that the relative error in the scaling relation \eqref{Eq:Nu} increases with smaller dimensions. For both the $(7,5,0)$ and $(5,3,2)$ truncations, we obtain a $15\%$ error in $d = 5$ dimensions, a $60 - 70\%$ error in $d = 4$ dimensions etc. This is an indication that the considered series expansions are not fully converged yet. Nevertheless, since we expect these scaling relations to hold for high enough orders, we employ Eq.\ \eqref{Eq:Nu} in the following to determine the exponent $\nu$, keeping in mind that the corresponding estimates will be associated with an error that is likely to decrease only when higher-order truncations are considered. In particular, the above numbers suggest that to reach a given precision, one will need to account for an increasing number of operators in $\Gamma_k$ in lower dimensions.

\begin{figure}[!t]
  \centering
  \includegraphics[width=0.46\textwidth]{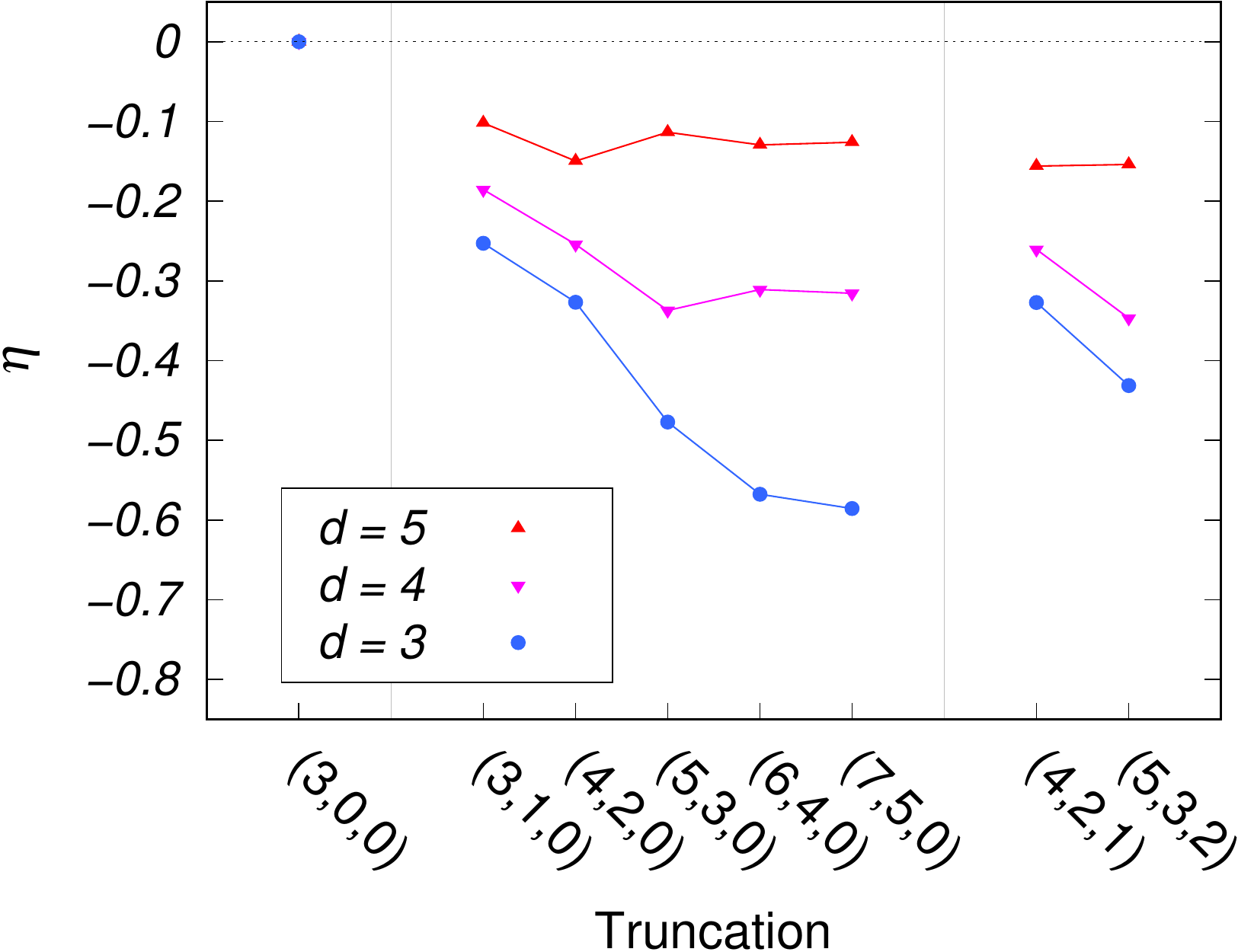}
  \caption{\label{Fig:CriticalExponentsTruncations}Anomalous dimension $\eta$ for different truncations of the scale-dependent effective action in $d = 3$, $4$, and $5$ dimensions.}
\end{figure}
\begin{table}[!t]
  \setlength{\tabcolsep}{2pt}  
  \def\arraystretch{1.5}
  \centering
  \begin{tabular}{cc|cS[table-format=-1.5,table-space-text-post="  "]S[table-format=-1.5,table-space-text-post="  "]S[table-format=-1.5,table-space-text-post="  "]}     \hline\hline
    \multicolumn{2}{c}{Critical exponent} && \multicolumn{1}{c}{$d = 3$} & \multicolumn{1}{c}{$d = 4$} & \multicolumn{1}{c}{$d = 5$} \\ \hline\hline
    \multicolumn{2}{c}{$\eta$} && -0.586(29) & -0.316(16) & -0.126(6) \\
    \multicolumn{2}{c}{$\sigma$} && 0.0742(56) & 0.2667(32) & 0.4033(12) \\
    \multicolumn{2}{c}{$\nu_c$} && 0.3581(19) & 0.3167(8) & 0.2807(2) \\
    \hline\hline
  \end{tabular}
  \caption{\label{Tab:CriticalExponents}Numerical values for the anomalous dimension $\eta$ and critical exponents $\sigma$, $\nu_c$ in $d = 3$, $4$, and $5$ dimensions. Here, we show our best estimates with errors to account for possible systematic effects (see Sec.\ \ref{Sec:Residual regulator dependence and principle of minimal sensitivity}). These values were obtained with an exponential regulator ($\alpha = 1$) [cf.\ Eq.\ \eqref{Eq:ExponentialRegulator}] and the truncation of the type $(7,5,0)$.}
\end{table}

\begin{table*}[!t]
  \setlength{\tabcolsep}{13pt}
  \def\arraystretch{1.5}
  \centering
  \begin{tabular}{c|S[table-format=1.6]|S[table-format=1.4]S[table-format=1.4]S[table-format=1.5]S[table-format=1.6]S[table-format=1.5]S[table-format=1.5]}
    \hline\hline
     Dimension & \mbox{functional RG} & \multicolumn{2}{c}{Ref.\ \cite{Gracey:2015tta}} & \multicolumn{1}{c}{Ref.\ \cite{Butera:2012tq}} & \multicolumn{1}{c}{Ref.\ \cite{Lai:1995}} & \multicolumn{1}{c}{Ref.\ \cite{Hsu:2005}} & \multicolumn{1}{c}{Ref.\ \cite{Gliozzi:2014jsa}} \\ \hline\hline
     $d = 3$ & 0.0742(56) & 0.0785 & 0.0747 & 0.076(2) & 0.0877(25) & 0.080(7) & 0.085(1) \\
     $d = 4$ & 0.2667(32) & 0.2616 & 0.2584 & 0.258(5) & 0.2648(15) & 0.261(12) & 0.2685(1) \\
     $d = 5$ & 0.4033(12) & 0.3989 & 0.3981 & 0.401(9) & 0.402(5) & 0.40(2) & 0.4105(5) \\ \hline\hline
    \end{tabular}
    \caption{\label{Tab:CriticalExponentsListResults}Different estimates for the critical exponent $\sigma$ (as compiled in Ref.\ \cite{Gracey:2015tta}) including results from the constrained three- and four-loop $\epsilon$ expansion \cite{Gracey:2015tta}, strong-coupling expansion \cite{Butera:2012tq}, Monte Carlo methods \cite{Lai:1995,Hsu:2005}, and conformal bootstrap \cite{Gliozzi:2014jsa}. The values obtained from the functional RG, with an exponential regulator ($\alpha = 1$) and truncation of the type $(7,5,0)$, lie within error bars of Refs.\ \cite{Butera:2012tq,Lai:1995,Hsu:2005}, and are slightly larger the values provided by constrained Pad\'e approximants of three- and four-loop $\epsilon$ expansion results \cite{Gracey:2015tta}, but are smaller than those obtained by conformal bootstrap methods \cite{Gliozzi:2014jsa}.}
\end{table*}

The scaling properties of the Yang-Lee edge are completely determined by the anomalous dimension $\eta$. Therefore, we may use Eqs.\ \eqref{Eq:NuC} and \eqref{Eq:Sigma} to calculate the critical exponents $\nu_c$ and $\sigma$. Our results are summarized in \mbox{Fig.\ \ref{Fig:CriticalExponents}} where we show the overall performance of different truncations in the range $3 \leq d \leq 6$ at the example of $\sigma$ and $\nu_c$, contrasted against the one- and two-loop $\epsilon$ expansion. In Fig.\ \ref{Fig:CriticalExponentsTruncations} we show the values for $\eta$ in $d = 3$, $4$, and $5$ dimensions for all truncations employed in this work, and our best estimates for the critical exponents $\eta$, $\nu_c$, and $\sigma$ are reported in Tab.\ \ref{Tab:CriticalExponents}. These values were obtained with the $(7,5,0)$ truncation for which, in contrast to the $(5,3,2)$ truncation, the values of $\eta$ seem to be reasonably close to their asymptotic values that are reached in the infinite $n_U$ and $n_Z$ limit [cf.\ Fig.\ \ref{Fig:CriticalExponentsTruncations}]. That is, we observe that larger orders of the finite field expansion are necessary to reach the asymptotic scaling exponents and it seems that this order increases for dimensions well below the upper critical dimension $d_c = 6$, which is consistent with our previous observation on the validity of scaling relations.

In Tab.\ \ref{Tab:CriticalExponents} and \ref{Tab:CriticalExponentsListResults} we account for a systematic bias due to our choice of the IR regulator (see Sec.\ \ref{Sec:Residual regulator dependence and principle of minimal sensitivity} for an in depth discussion of this issue). We remark that the difference in the values of the anomalous dimension between different high-order truncations is typically larger than that obtained for the critical exponents $\sigma$ and $\nu_c$, which is reflected in the errors for these quantities (cf.\ Tab.\ \ref{Tab:CriticalExponents}). This effect has also been observed with other methods and may be attributed to the scaling relations \eqref{Eq:NuC} and \eqref{Eq:Sigma} that yield a smaller error for the exponents $\nu_c$ and $\sigma$  (see, e.g., Ref. \cite{Gracey:2015tta}).

Comparing our estimates for the critical exponent $\sigma$ to a recent
compilation of available data on the Yang-Lee edge critical scaling exponents provided in Ref.\ \cite{Gracey:2015tta}, cf.\ Tab.\ \ref{Tab:CriticalExponentsListResults}, we find that our values lie within the error bounds provided by other methods, e.g., Refs.\ \cite{Butera:2012tq,Lai:1995,Hsu:2005}. They lie slightly above the values obtained from constrained Pad\'e approximants of three- and four-loop $\epsilon$ expansion results \cite{Gracey:2015tta}, but are in general smaller than those values obtained from a recent conformal bootstrap analysis \cite{Gliozzi:2014jsa}. Considering the fact, that our numerical implementation of the RG flow equations is not overly sophisticated (limiting the truncations that can be considered to a relatively small number of operators) it is quite remarkable that our present results are competitive with other data in the literature.

\section{Relevance of composite operators and quality of finite truncations}
\label{Sec:Relevance of composite operators and quality of finite truncations}

We observe that certain truncations of the scale-dependent effective
action, of the type $(n_U, 0, 0)$, $n_U > 3$, are inadequate to
investigate the Yang-Lee scaling behavior. In fact, for these
truncations, the Yang-Lee fixed point is unstable below $d \approx
5.6$.\footnote{We remark that this observation depends on the choice
  of the IR regulator. While the Lee-Yang edge fixed point is unstable for the smooth exponential regulator \eqref{Eq:ExponentialRegulator} ($\alpha = 1$), this is not the case for the optimized Litim regulator \cite{Litim:2000ci,Litim:2001up}. However, the latter is not immediately applicable at higher orders in the derivative expansion.} This is certainly surprising and in conflict with other available data \cite{Butera:2012tq,Lai:1995,Hsu:2005,Gliozzi:2014jsa,Gracey:2015tta}. However, this behavior can be understood by examining the effect of operator insertions at the level of the one-loop $\epsilon = 6-d$ expansion, as considered in Refs.\ \cite{Amit:1977,Kirkham:1979}.

In particular, we consider the renormalization of quartic operators at the Yang-Lee fixed point. This requires the simultaneous renormalization of all operators that carry the same canonical dimension as $\delta\phi_k^4$, which mix under renormalization \cite{Brezin:1976}. In $d = 6 -\epsilon$ dimensions these operators can be listed as follows (up to total derivative contributions)
\begin{subequations}
  \begin{eqnarray}
    A_{1,k} &=& \delta\phi_k^4 /4! , \label{Eq:Operator1} \\
    A_{2,k} &=& k^{\epsilon/2} \delta\phi_k \left( \partial \delta\phi_k \right)^2 / 2 , \\
    A_{3,k} &=& k^{\epsilon} \left( \Box \hspace{1pt} \delta\phi_k \right)^2 / 2 . \label{Eq:Operator3}
  \end{eqnarray}
\end{subequations}
Note that they simply correspond to particular contributions in the finite series expansion of $U_k(\phi)$, $Z_k(\phi) (\partial \phi)^2$, and $W_k(\phi) \left(\Box \phi\right)^2$, respectively, around the homogeneous field expectation value $\bar{\phi}_k$. Different truncations of the scale-dependent effective action are distinguished by either including or neglecting some of these operators, \eqref{Eq:Operator1} -- \eqref{Eq:Operator3}. The $(n_U, 0, 0)$-type truncations, for instance do not include operators $A_{2,k}$ and $A_{3,k}$, while truncations of the type $(n_U, n_Z, 0)$ do not include $A_{3,k}$.

Treating the operators \eqref{Eq:Operator1} -- \eqref{Eq:Operator3} on an equal footing, both $A_{2,k}$ and $A_{3,k}$ turn out to be more relevant in $d < 6$ dimensions than the quartic interaction $A_{1,k}$. Indeed, from a one-loop calculation \cite{Amit:1977,Kirkham:1979}, we obtain the following eigenvalues of the stability matrix: $\lambda_4 = -2$, $\lambda_5 = -2 - \epsilon/9$, and $\lambda_6 = -2 - 19 \epsilon / 9$. Each of them corresponds to a different linear combination of operators \eqref{Eq:Operator1} -- \eqref{Eq:Operator3}. One can show that the dominant contribution to $\lambda_4$ comes from $A_{3,k}$, for $\lambda_5$ it is the operator $A_{2,k}$, and for $\lambda_6$ it is $A_{1,k}$ that contributes the
most. Thus, one might conclude that any truncation that includes only the quartic interaction $A_{1,k}$ is ill-defined, as it neglects the more relevant contributions, namely $A_{2,k}$ and $A_{3,k}$. Interestingly, it is sufficient to consider truncations of the type $(n_U, n_Z, 0)$ to stabilize the Yang-Lee edge fixed point. While $(n_U, 0, 0)$-type truncations, $n_U > 3$, fail to produce a Yang-Lee edge fixed point below $d\approx 5.6$, the $(n_U, n_Z, 0)$ truncations allow us to identify the corresponding scaling solution all the way down to $d = 3$ [cf.\ Fig.\ \ref{Fig:CriticalExponents}]. In general, we expect that the scale-dependent effective action needs to respect the properties of the theory under simultaneous renormalization of operators with the same canonical dimension. This is important to define consistent truncations that are adequate to describe the Yang-Lee edge critical point.

\section{Residual regulator dependence and principle of minimal sensitivity}
\label{Sec:Residual regulator dependence and principle of minimal sensitivity}

To determine the critical scaling properties of a given model, we may in principle choose any regulator function $R_k = R_k(q)$ as long as it satisfies the appropriate limiting behavior $\lim_{k\rightarrow 0} R_k = 0$ and $\lim_{\Lambda \rightarrow \infty} R_{k = \Lambda} = \infty$. Indeed, if an exact solution to the functional flow equation for $\Gamma_k$ were available, the calculated observables should not depend on the way we choose to regularize the theory in the IR and therefore must be independent of the regulator. However, in practice, we are bound to consider truncations of the coupled infinite set of flow equations. This yields a finite set of RG equations for which one observes a residual regulator dependence \cite{Pawlowski:2015mlf}. To investigate this effect, we define a one-parameter family of functions
\begin{equation}
  R_{\alpha, k} = \alpha R_k , 
\end{equation}
with $\alpha > 0$, and consider the $\alpha$ dependence of the critical exponents. We employ the following set of exponential regulators
\begin{equation}
  R_{\alpha, k}^{\textrm{exp}} = \frac{\alpha \bar{Z}_k q^2}{\exp(q^2 / k^2) - 1} ,
  \label{Eq:ExponentialRegulator}
\end{equation}
for this analysis.\footnote{Note that the regulator should be sufficiently smooth in momentum space if higher order approximations in the derivative expansion are considered (see, e.g., Ref.\ \cite{Canet:2003qd}).} One may identify an optimal value of $\alpha$, which is determined by the principle of minimum sensitivity \cite{Canet:2003qd}. It states that the value of any given observable that is least sensitive to changes in $\alpha$ can be considered the best estimate for that quantity. Since by virtue of scaling relations all critical exponents at the Yang-Lee edge critical point can be expressed in terms of the anomalous dimension $\eta$, we apply this criterion to $\eta = \eta(\alpha)$, i.e., to find the optimal value, we require that
\begin{equation}
  \eta'(\alpha = \alpha_{\textrm{opt}}) = 0 .
\end{equation}
In Tab.\ \ref{Tab:RegulatorDependence} we compare the values of $\eta(\alpha)$ evaluated for $\alpha = 1$ as well as $\alpha = \alpha_{\textrm{opt}}$ in different dimensions and determine the relative error $\Delta\eta / \eta(\alpha_{\textrm{opt}}) \equiv \left[\eta(1) - \eta(\alpha_{\textrm{opt}})\right] / \eta(\alpha_{\textrm{opt}})$. Largely independent of dimension, the anomalous dimension evaluated at $\alpha = 1$ seems to be slightly overestimated with a relative error of approximately $3\%$. From this comparison we conclude that $\eta(\alpha = 1)$ is typically already a good approximation to the optimal value $\eta(\alpha_{\textrm{opt}})$.

\begin{table}[!t]
  \setlength{\tabcolsep}{5pt}
  \def\arraystretch{1.5}
  \centering
  \begin{tabular}{ccS[table-format=-1.4,table-space-text-post=""]S[table-format=-1.4,table-space-text-post=""]S[table-format=-1.4,table-space-text-post=""]}
    \hline\hline
    \multicolumn{1}{c}{} && \multicolumn{1}{c}{$d = 3$} & \multicolumn{1}{c}{$d = 4$} & \multicolumn{1}{c}{$d = 5$} \\ \hline
    $\eta(\alpha = 1)$ && -0.3270 & -0.2542  & -0.1498 \\
    $\eta(\alpha = \alpha_\text{opt})$  && -0.3340 & -0.2587 & -0.1500 \\
    Relative error && \multicolumn{1}{c}{2.1\%} & \multicolumn{1}{c}{1.7\%} & \multicolumn{1}{c}{1.3\%} \\
    \hline\hline
  \end{tabular}
  \caption{\label{Tab:RegulatorDependence}Anomalous dimension $\eta =
    \eta(\alpha)$ at the Yang-Lee edge critical point in $d$ dimensions, evaluated for the (deformed) exponential regulator $R_{\alpha, k}^{\textrm{exp}}(q) = \alpha \bar{Z}_k q^2 \left[\exp (q^2 / k^2) -1 \right]^{-1}$ with $\alpha > 0$. The optimal value of $\alpha$ depends on the dimension, i.e., $\alpha_{\textrm{opt}} = \alpha_{\textrm{opt}}(d)$ [cf.\ Fig.\ \ref{Fig:AnomalousDimensionRegulatorDependence}]. The shown values were obtained using a truncation of the scale-dependent effective action $\Gamma_k$ defined by the index set $(4,2,0)$.}
\end{table}

\begin{figure}[!t]
  \centering
  \includegraphics[width=0.46\textwidth]{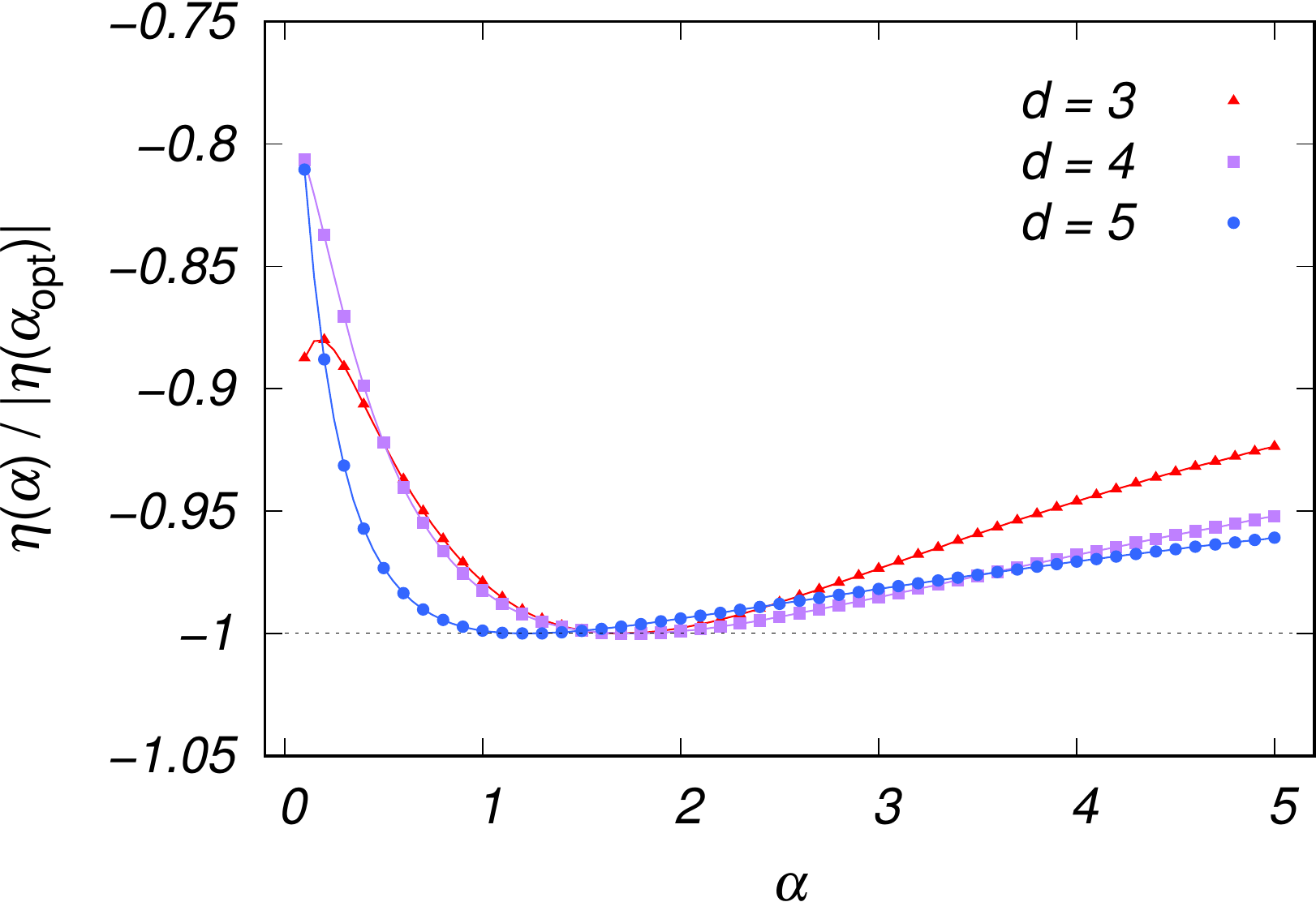}
  \caption{\label{Fig:AnomalousDimensionRegulatorDependence}Rescaled anomalous dimension $\eta(\alpha) / |\eta(\alpha_{\textrm{opt}})|$ shown as a function of $\alpha$. Different curves correspond to data obtained in $d = 3$, $4$, and $5$ dimensions, respectively. The optimal value $\alpha_{\textrm{opt}}$ for which the critical exponent is least sensitive to changes in the deformation parameter, i.e., \mbox{$\eta'(\alpha = \alpha_{\textrm{opt}}) = 0$}, shifts to larger values as the dimension $d$ is lowered. The displayed values were obtained for a truncation of the scale-dependent effective action $\Gamma_k$ of the type $(4,2,0)$.}
\end{figure}

In principle, $\alpha_{\textrm{opt}}$ might depend on the dimension. Indeed, as shown in Fig.\ \ref{Fig:AnomalousDimensionRegulatorDependence}, the optimal value of $\alpha$ shifts to larger values  when the dimension $d$ is lowered and eventually stabilizes around $\alpha\approx 1.7$. Although the value of $\alpha_{\textrm{opt}}$ increases, the relative error in $\eta$ remains roughly constant. At this point, we remark that below $d = 4$ an ambiguity appears: $\eta(\alpha)$ develops a second extremum, a local maximum, for $\alpha < 1$ [cf.\ Fig.\ \ref{Fig:AnomalousDimensionRegulatorDependence}]. However, we do not consider this solution to be physical and define $\alpha_{\textrm{opt}}(d)$ as the analytically continued local minimum from $d = 6 - \epsilon$.

Since the search for fixed points of the RG $\beta$ functions becomes quite demanding numerically for higher-order truncations, we use this information to limit our calculations to the case $\alpha = 1$ and estimate the corresponding systematic error in $\eta(\alpha = 1)$ at the $3 - 5\%$ level (within the considered one-parameter family of regulators). This systematic effect in the estimation of the anomalous dimension has been accounted for and is indicated explicitly as a systematic error in the summary of our results in Tab.\ \ref{Tab:CriticalExponents} and \ref{Tab:CriticalExponentsListResults}.

\section{Conclusions}
\label{Sec:Conclusions}

In this work we have examined the critical scaling properties of the
Yang-Lee edge, or $\phi^3$, theory in dimensions $3 \leq d \leq 6$. We find our results in good agreement with available data in the literature, which includes high-temperature series expansions, results from the $\epsilon$ expansion, strong coupling expansion, and Monte Carlo methods. While our results are consistent with the strong coupling expansion and Monte Carlo methods Refs.\ \cite{Butera:2012tq,Lai:1995,Hsu:2005}, our estimates for the critical exponent $\sigma$ are slightly larger than the values obtained from constrained Pad\'e approximants for three- and four-loop $\epsilon$ expansion results \cite{Gracey:2015tta}, and generally lie below those from a conformal bootstrap analysis \cite{Gliozzi:2014jsa}. We expect that truncations at higher orders in the derivative and field expansion will improve our estimates for the critical exponents. However, more elaborate numerical treatment is necessary to study such truncations.

We have shown that the stability of nontrivial fixed point associated to the Yang-Lee edge singularity is sensitive to the insertion of operators that mix under renormalization. This might seem surprising since a similar behavior is not observed in applications of the functional RG to establish the scaling behavior at the Ising critical point. However, comparing our results with a stability analysis at the fixed point to one-loop order in the $\epsilon = 6-d$ expansion provides a qualitative explanation for the observed lack of stability of the Yang-Lee edge fixed point for $(n_U,0,0)$-type truncations ($n_U > 3$) of the scale-dependent effective action.

Finally, we remark on possible applications of this work. Based on mean-field arguments, one expects another thermodynamic singularity in the low-temperature phase of the Ising model ($T < T_c$) with exactly the same critical exponents as those of the Yang-Lee edge point -- the spinodal singularity. The corresponding critical point appears on the metastable branch of the free energy and is usually associated with the classical limit of metastability. However, its existence (beyond mean-field) as well as its scaling properties have been subject to some debate \cite{Saito:1978,Gunton:1978,Klein:1981,Unger:1984}. It would be interesting to understand the relation between the Yang-Lee edge point and the spinodal singularity~\cite{Privman:1982,Fonseca:2001dc}. We intend to address these issues in a future publication.

\begin{acknowledgements}
  We thank B.\ Delamotte for valuable insights and for helpful advice regarding the numerical implementation of the RG flow equations. This research is funded by the European Research Council under the European Union’s Seventh Framework Programme FP7/2007-2013, 339220. This material is based upon work supported by the U.S. Department of Energy, Office of Science, Office of Nuclear Physics under Award Number DE-FG0201ER41195.
\end{acknowledgements}

\bibliography{references}

\end{document}